\documentclass{ptephy_v1}

\usepackage{amsmath,amssymb,amsthm, graphicx, latexsym, eepic}

\newcommand{\beq}{\begin{equation}}
\newcommand{\eeq}{\end{equation}}
\newcommand{\beqa}{\begin{eqnarray}}
\newcommand{\eeqa}{\end{eqnarray}}
\newcommand{\CR}{\nonumber \\}

\newcommand{\trace}{{\rm Tr}}
\newcommand{\gq}{\mathfrak{q}}

\numberwithin{equation}{section}

\begin{document}

\title{Quiver matrix model of ADHM  type and \\ BPS state counting 
in diverse dimensions}


\author{Hiroaki Kanno}
\affil{Graduate School of Mathematics and KMI, \\
Nagoya University, Nagoya, 464-8602, Japan \email{kanno@math.nagoya-u.ac.jp}}
%
%
%

\begin{abstract}%
We review the problem of BPS state counting described by the generalized quiver matrix model
of ADHM type. In four dimensions the generating function of the counting gives the Nekrasov partition function
and we obtain generalization in higher dimensions. By the localization theorem, the partition function 
is given by the sum of contributions from the fixed points of the torus action, which are labeled by
partitions, plane partitions and solid partitions. The measure or the Boltzmann weight 
of the path integral can take the form of the plethystic exponential. Remarkably 
after integration the partition function or the vacuum expectation value is again expressed 
in plethystic form. We regard it as a characteristic property of the BPS state 
counting problem, which is closely related to the integrability. 
\end{abstract}

\subjectindex{A10 Integrable systems and exact solutions, 
B27 Topological field theory}

\maketitle

%
%

\section{Introduction}

It is well known the instantons (anti-self-dual connections) in four dimensional gauge theory allow
ADHM construction \cite{Atiyah:1978ri},\cite{Corrigan:1983sv}. From the viewpoint of string theory 
the ADHM description can be obtained by considering $D4$-$D0$ system in type IIA string theory, 
where the matrices, which are basic dynamical variables in ADHM construction,
come from the open strings connecting $D$-branes\footnote{In this article we only consider $U(n)$ gauge theory.}
\cite{Witten:1994tz},\cite{Witten:1995gx},\cite{Douglas:1995bn},\cite{Douglas:1996uz}. 
The low energy effective theory on the world volume of $D$-branes is the dimensional reduction of ten dimensional
super Yang-Mills theory. The original anti-self-duality of the gauge field is translated to the BPS condition for the world volume 
theory on $D4$-branes, while the ADHM equations are obtained as the BPS condition on $D0$-branes.
Since the world volume of $D0$-branes has no spacial direction, the theory is reduced to supersymmetric quantum mechanics
(in fact matrix model), which we call ADHM matrix model.

The ADHM description of the four dimensional instantons (or the BPS solitons in five dimensional theory
from the viewpoint of $M$-theory) also plays a significant role in the computation of the instanton partition 
function of Nekrasov \cite{Nekrasov:2002qd},\cite{Losev:2003py},\cite{Nekrasov:2003rj}, 
which provides a microscopic derivation of the Seiberg-Witten prepotential of 
four dimensional $\mathcal{N}=2$ Yang-Mills theory. By introducing sufficiently large number of torus action
on the ADHM moduli space, we can employ the Atiyah-Bott type localization formula to compute the path integral.
The fixed points of the topic action are isolated and labeled by a tuple of partitions (or Young diagrams).
Then the partition function is expressed as a summation over the contribution from each fixed point,
which is in turn given by the equivariant character of the tangent space at the fixed point as a module of the torus
action.

In the following we will argue some of intriguing aspects in generalizing this story 
to higher dimensions by replacing $D4$-brames with $Dp~(p=2d=6,8)$-branes, where 
the fixed points are labeled by higher dimensional generalizations of 
the partition, called plane partition ($d=3$) and solid partition ($d=4$). 
The BPS condition on $D6$ and $D8$-branes can be identified with the higher dimensional instanton equation 
in six and eight dimensions, respectively \cite{Baulieu:1997jx}, 
while the BPS condition on $D0$-branes gives what we call quiver matrix model of ADHM type. 
In the same manner as the four dimensional case, the moduli space $\mathcal{M}_{n,k}$ 
of the quiver matrix model is topologically labeled by the number $n$ of $Dp$-branes and
the number $k$ of $D0$-branes. We call $k$ instanton number 
in analogy with the four dimensional case.
To obtain the partition function of $U(n)$ gauge theory on $Dp$-brane, 
we will fix $n$ and take a summation of $k$ over non-negative integers.

\subsection{Fixed points of the torus action and $(d-1)$-partitions}
Let $t_i$ collectively denote equivariant parameters of the torus action on the moduli space $\mathcal{M}_{n,k}$
of matrix equations of ADHM type. In general, they consist of the equivariant parameters of the torus action on 
the (flat) space-time coordinate $(z_1, \cdots, z_d) \in \mathbb{C}^d$ (the $\Omega$ background parameters of Nekrasov),
the Cartan subgroup of the gauge symmetry $G_{C}=U(n)$ (the Coulomb moduli parameters)
and of the flavor symmetry $G_{F}$ (mass parameters). 
We can identify the equivariant $K$ group of a point with the ring of Laurant polynomials in 
the equivarinat parameters $K_T(\mathrm{pt}) = \mathbb{C} [ t_i^{\pm 1}]$.
Hence the equivariant character at the fixed points takes the value in $K_T(\mathrm{pt})$.


\begin{figure}[h]
 \begin{center}
 \begin{picture}(100,120)
 \put(-150,0){\includegraphics[width=4cm,clip]{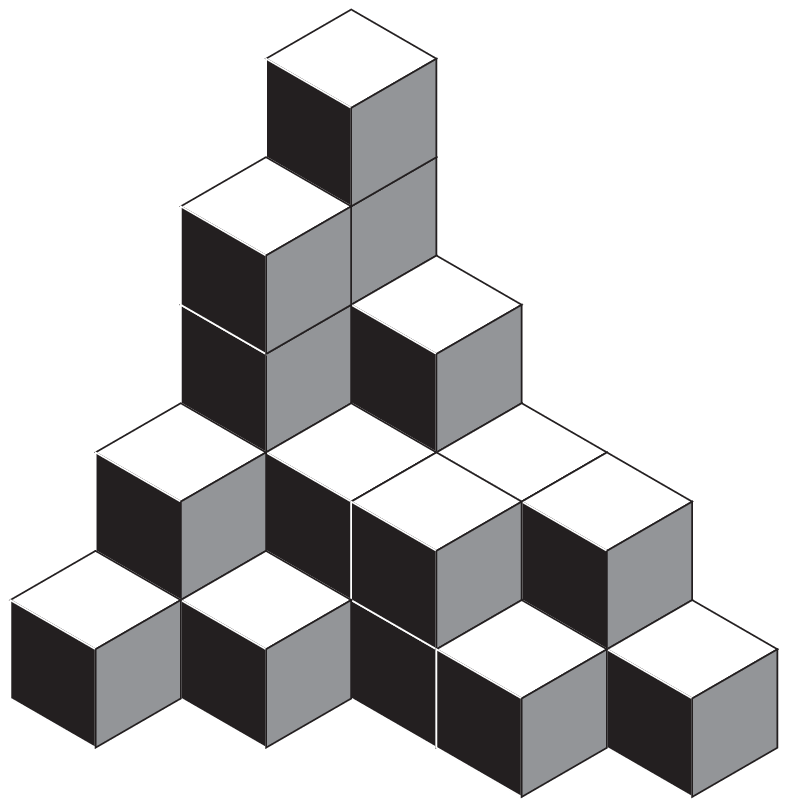}}
 \put(0, 50){$\pi = \left(\begin{array}{ccccc}
 5 & 3 & 2 & 2 &1 \\
 4 & 2 & 2 & 1 &  \\
 2 & 1 &  &  & \\
  1 &  &  &  & \\
 \end{array}\right)$,}
  \put(150, 50){$|\pi| = 26$.}
  \end{picture}
 \end{center}
 \caption{Plane partition as 3 dimensional Young diagram}
 \label{fig:3DYoung}
\end{figure}


Recall that a partition $\lambda = (\lambda_1 \geq \lambda_2 \geq \cdots \geq \lambda_\ell >0)$ is a non-increasing
sequence of positive integers such that $\lambda_{\ell+1}=0$ for finite $\ell$. It is useful to represent $\lambda$ 
in terms of the Young diagram. We denote $|\lambda| = \sum_{i=1}^\infty \lambda_i$, which is the total number of 
boxes (cells) in the corresponding Young diagram. We can consider higher dimensional generalization or
a $(d-1)$-partition $\pi = \{ \pi_{i_1, \cdots, i_{d-1} } \}~(i_1, \cdots, i_{d-1}) \in \mathbb{Z}_{>0}^{d-1}$,
which is an array of positive integers with (obvious) higher dimensional generalization of the non-increasing 
condition; for example $ \pi_{i, j} \geq  \pi_{i+1, j},~ \pi_{i, j} \geq  \pi_{i, j+1}$ when $d=3$
and $\pi_{i_1, \cdots, i_{d-1}} \neq 0$ for only finite set of $(i_1, \cdots, i_{d-1})$
 (see Fig.\ref{fig:3DYoung}). When $d=3$ and $d=4$, it is usually called plane and solid partition, respectively. 
We define $|\pi| = \sum_{(i_1, \cdots, i_{d-1})}  \pi_{i_1, \cdots, i_{d-1} }$, which means 
the volume (the number of boxes, cubes $\cdots$) of the $(d-1)$-partition $\pi$.
It turns out that the fixed points of the toric action on $\mathcal{M}_{n,k}$ are isolated 
and in one to one correspondence with the set of $n$-tuples of $(d-1)$-partitions 
$\vec{\pi} = (\pi^\alpha)_{\alpha=1}^n$, and that
$|\vec{\pi}| := \sum_{\alpha=1}^n |\pi^\alpha |$ is identified with the instanton number $k$. 
Thus in the sector of instanton number $k$, we can reduce the quiver matrix model 
to a statistical model with the configuration space $\Pi_k^n := \{ \vec{\pi} \vert |\vec{\pi}|=k \}$,
where the equivariant character at each fixed point $\vec{\pi}$ gives the Boltzmann weight of the model.

\subsection{Partition function and plethystic exponential}

It is interesting that the Boltzmann weight derived from the ADHM matrix model 
takes the form of the plethystic exponential (see section 2 for definition) $\hbox {P.E.} [\chi_{\vec{\pi}}(t_i)]$.
We define the topological partition function by a weighted sum over the total
configuration space  $\cup_{k \geq 0}\Pi_k^n$, where introducing the box counting parameter $\gq$,
we multiply the volume (the number of boxes, cubes $\cdots$) of
 $(d-1)$-partitions $|\pi|$ as the additional Boltzman weight;
\beq\label{top}
Z_{\rm {top}} (t_i; \gq) := \big\langle \hbox {P.E.} [ \chi_{\vec{\pi}}(t_i) ] \big\rangle = \sum_{\pi} \gq^{|\pi|}  
\hbox {P.E.} [\chi_{\vec{\pi}}(t_i)].
\eeq
Namely if we identify the instanton number $k$ as the particle number,
the topological partition function corresponds to the grand canonical ensemble in statistical mechanics.
The phenomena on which we will focus in this article is that in the computation of the topological partition function, 
the expectation value of the plethystic exponential is again expressed by the plethystic exponential;
\beq\label{super}
 \big\langle \hbox {P.E.} [ \chi_{\pi}(t_i) ] \big\rangle = \hbox {P.E.} [ F(t_i ;\gq) ]. 
\eeq
Since the plethystic exponential can be regarded as the character of the symmetric algebra $S^{\bullet} V$ 
of a $G$-module $V$, this is an example of \lq\lq super\rq\rq-integrability
that the expectation value of the character gives another character, which we encounter 
typically in matrix model and plays an important role for extending the realm of symmetric functions
\cite{Morozov:2018fjb},\cite{Mironov:2019uoy}.

When $d=3$ with the computation of the topological partition function $Z_{\rm {top}} (t_i; \gq)$
we may associate equivariant (or $K$-theory) vertices \cite{Nekrasov:2014nea}, 
which are generalizations of the refined topological vertex \cite{Awata:2005fa},\cite{Iqbal:2007ii}. 
In fact in an appropriate limit of the $\Omega$ background parameters $q_i$,
the equivariant vertex reduces to the refined topological vertex. 
Since the refined topological vertex is characterized as the intertwining operator of the quantum toroidal algebra
of $\mathfrak{gl}(1)$ \cite{Awata:2011ce}, it is tempting to expect some quantum algebras
behind the  \lq\lq super\rq\rq-integrability \eqref{super}.
Furthermore, since physically the partition function \eqref{top} is nothing but the generation function of 
the numbers of BPS states, this seems to be along the same line of BPS/VOA correspondence,
the correspondence of the algebra of BPS states with the chiral algebra of some 2 dimensional CFT.
It may be interesting to look at cohomological Hall algebra associated with the quiver of ADHM type
\cite{Rapcak:2018nsl}. 

The paper is organized as follows;
In the next section we introduce the plethysitc exponential. 
We can regard it as the character of the symmetric algebra and hence
it plays a significant role in this article. 
After presenting ADHM type matrix model equations coming from the BPS condition of 
$D$-brane system in section 3, we discuss a matrix model formulation 
or the measure for eigenvalues of matrices in section 4.
The measure is given in terms of the plethystic exponential and hence
naturally expressed by the power sum functions of eigenvalues. 
In section 5 we compute the equivariant character of the tangent space at 
the fixed points. Finally we present the plethysitic forms of the partition function inspection 6.
In each section after section 4, we first review the well-established case of $d=2$
 (the original ADHM equation) and then try to generalize it to higher
 dimensions. From the view point of mathematics, one of the crucial points
 is that though the fixed points of the torus action are still isolated 
 and labeled by higher dimensional generalization of the partition, 
 the tangent space at each fixed point is not smooth anymore and it is defined only virtually. 
 
\bigskip\bigskip

I would like to dedicate this article to the memory of Prof. Tohru Eguchi who passed away last year.
My collaboration with him started when both of us participated the inaugural project at Newton Institute
in summer of 1992. Our interest was an interplay of topological string as two dimensional TQFT 
and integrable systems such as $w_{1+\infty}$ algebra, 
Toda lattice hierarchy \cite{Eguchi:1992tp},\cite{Eguchi:1993xx},\cite{Eguchi:1994np},
which became one of main themes in my research afterwards. 
After almost a decade I had a second chance of collaboration 
on five dimensional lift of Seiberg-Witten theory, Nekrasov partition function
and topological strings \cite{Eguchi:2000fv},\cite{Eguchi:2003sj},\cite{Eguchi:2003it},
which are closely related to the subject reviewed in the present paper.
I am very grateful to Eguchi-san for these fruitful and inspiring collaborations.
Though I was not his student,  I learned how to enjoy the research through the collaboration 
with Eguchi-san.

\bigskip\bigskip


\section{Plethystic exponential}

For a function $F(t_1, t_2, \cdots, t_\ell)$ we define the plethystic exponential by
\beq
\hbox {P.E.} [ F(t_1, t_2, \cdots,t_\ell)] = \exp
\left( \sum_{k=1}^\infty \frac{1}{k} F(t_1^k, t_2^k, \cdots, t_\ell^k) \right).
\eeq
Let us assume that $F(t_1, t_2, \cdots, t_\ell)$ can be expanded as follows;
\beq
F(t_1, t_2, \cdots, t_\ell) = \sum_{n_1, \cdots, n_\ell \in \mathbb{Z}} a_{n_1 \cdots n_\ell} t_1^{n_1} \cdots t_\ell^{n_\ell}
\eeq
with $a_{0 \cdots 0} =0$. Then we see
\beqa
\sum_{k=1}^\infty \frac{1}{k} F(t_1^k, t_2^k, \cdots, t_\ell^k) 
&=& \sum_{n_1, \cdots, n_\ell \in \mathbb{Z}}  a_{n_1 \cdots n_\ell} \sum_{k=1}^\infty \frac{1}{k} t_1^{kn_1} \cdots t_\ell^{kn_\ell} \CR
&=&  - \sum_{n_1, \cdots, n_\ell \in \mathbb{Z}} a_{n_1 \cdots n_\ell}  \log ( 1-  t_1^{n_1} \cdots t_\ell^{n_\ell}).
\eeqa
Thus the plethystic exponential factorizes as an infinite product;
\beq
\hbox {P.E.} [ F(t_1, t_2, \cdots,t_\ell)]  
= \prod_{n_1, \cdots, n_\ell \in \mathbb{Z}}  ( 1- t_1^{n_1} \cdots t_\ell^{n_\ell})^{-a_{n_1 \cdots n_\ell}} .
\eeq
In fact when $F(t_1, t_2, \cdots, t_\ell)$ is a character of a $G$ module $V$,
with $t_i$ parametrizing the Cartan subgroup of $G$;
\beq
F(t_1, t_2, \cdots, t_\ell) = \trace_V~g,
\eeq
the plethystic exponential computes the character of the symmetric algebra $S^k V$;
\beq
\sum_{k=1}^\infty s^k~\trace_{S^kV} g^k = \hbox {P.E.} [ s \cdot F(t_1, t_2, \cdots,t_\ell)].
\eeq

The MacMahon function is a typical example of the plethystic exponential;
\beq
M(t) := \prod_{n=1}^{\infty} (1 - t^n)^{-n} = \exp \left( \sum_{k=1}^\infty \frac{1}{k [t^k]^2} \right),
\eeq
where we have introduced the notation
\beq
[x] := x^{\frac{1}{2}} - x^{-\frac{1}{2}} = -[x^{-1}].
\eeq
Note that
\beq
F(t) = \frac{1}{[t]^2} = t \frac{\partial}{\partial t} \left( \frac{1}{1-t} \right) = \sum_{n=1}^\infty n t^n.
\eeq
Another example which is also ubiquitous in our computation is
\beq
(x;q)_\infty = \prod_{n=0}^\infty ( 1- xq^n) =  \hbox{P.E.} \left[  - \frac{x}{1-q}   \right]
= \hbox{P.E.} \left[ \frac{x/\sqrt{q}}{[q]}   \right].
\eeq

It is curious that the generating function of the counting of solid partitions does not seem
to allow a plethystic expression. In fact the conjecture of MacMahon, which assumes a plethystic form,
fails.


\section{ADHM type equation as BPS condition}

To write down the matrix equations of ADHM type, 
we introduce two vector spaces $N$ and $K$ with $\dim_{\mathbb{C}} N =n$ and $\dim_{\mathbb{C}} K =k$.
They are associated with $Dp(p=2d=4,6,8)$ and $D0$-branes, respectively and the dimensions give the numbers of these branes. 
ADHM type equation is supposed to describe the BPS bound states of $D0$-branes (instantons) with the background $Dp$-branes.
In all the cases the equation of motion is invariant under the gauge symmetry $U(k)$ acting on the vector space $K$.
Note that since the matrix equations of ADHM type describes the theory on $D0$-branes 
the gauge symmetry is $U(k)$, while $U(n)$ symmetry on $Dp$-branes are regarded as the flavor symmetry.
In the following we list the equations of quiver matrix model.  There are two types of open string with boundary on $D0$ branes;
one is $k \times k$ matrix in $\mathrm{Hom}_{\mathbb{C}} (K,K)$, where both ends are attached to $D0$-branes
and the other is $k \times n$ matrix in $\mathrm{Hom}_{\mathbb{C}} (N,K)$ together with the conjugate
which describes open strings stretching between $D0$ and $Dp$-branes.

\begin{enumerate}
\item $d=2, X = \mathbb{C}^2$ ($D0$-$D4$ system, the original ADHM equation) \cite{Nakajima};
\beqa
\mu_{\mathbb{C}} &=& \left[ B_1, B_2 \right] + IJ = 0, \\
\mu_{\mathbb{R}}(\zeta )&=& \left[ B_1, B_1^\dagger \right] + \left[ B_2, B_2^\dagger \right] 
+ II^\dagger - J^\dagger J - \zeta\cdot E_{k \times k} =0 \quad (\zeta >0),
\eeqa
where $B_{1,2} \in \mathrm{Hom}_{\mathbb{C}} (K,K)$ and $I, J^\dagger \in \mathrm{Hom}_{\mathbb{C}} (N,K)$.
\item $d=3, X = \mathbb{C}^3$ ($D0$-$D6$ system)  \cite{Nekrasov:JJM},\cite{Jafferis:2007sg},\cite{Cirafici:2008sn};
\beqa
\mu_{\mathbb{C}} &=& \left[ B_i, B_j \right] + \frac{1}{2} \epsilon_{ijk} \left[ B_k^\dagger, Y \right] = 0, \\
\mu_{\mathbb{R}} (\zeta) &=& \sum_{i=1}^3  \left[ B_i, B_i^\dagger \right] + \left[ Y,  Y^\dagger\right] + I I^\dagger - \zeta\cdot E_{k \times k} =0 \quad (\zeta >0), \\
\mu_{B} &=& Y \cdot I =0,
\eeqa
where $B_{1,2,3}, Y \in \mathrm{Hom}_{\mathbb{C}} (K,K)$ and $I\in \mathrm{Hom}_{\mathbb{C}} (N,K)$.
\item $d=4, X = \mathbb{C}^4$ ($D0$-$D8$ system) \cite{Nekrasov:2017cih},\cite{Nekrasov:2018xsb};
\beqa
\mu_{\mathbb{C}} &=& \left[ B_a, B_b \right] + \frac{1}{2} \Omega_{abcd} \left[ B_c^\dagger, B_d^\dagger \right] = 0, \\
\mu_{\mathbb{R}} (\zeta) &=& \sum_{i=1}^4  \left[ B_i, B_i^\dagger \right] + I I^\dagger - \zeta\cdot E_{k \times k} =0 \quad (\zeta >0),
\eeqa
where $B_{1,2,3,4 } \in \mathrm{Hom}_{\mathbb{C}} (K,K)$, $I \in \mathrm{Hom}_{\mathbb{C}} (N,K)$
and $\Omega_{abcd}$ is the component of the Calabi-Yau $4$ form, with $\Omega \wedge \overline{\Omega} = \mathrm{vol}_8$.
\end{enumerate}

The origin of $Y$ in the case of $d=3$ is rather subtle.  But the equations can be obtained 
by a dimensional reduction of those for $d=4$ by putting $Y=B_4$.
Or we can regards it as a consequence of \lq\lq tachyon condensation\rq\rq\ \cite{Nekrasov:2018xsb}.
The additional condition $\mu_{B}=0$, which only appears for $d=3$, 
means that $D0$ branes cannot escape along the normal direction to $D6$-branes.
Similar condition appears in the BPS condition for the spiked 
instanton \cite{Nekrasov:2015wsu},\cite{Nekrasov:2016qym},\cite{Nekrasov:2016ydq},\cite{Nekrasov:2016gud}. 
Thus it might be more natural to consider $D8$-$D6$-$D0$ system as a generalization of the spiked instanton. 
 It has been argued that a constant $B$ field (a background flux) is required for the existence of 
bound states of $D0$-$D6$ and $D0$-$D8$ systems \cite{Witten:2000mf},\cite{Ohta:2001dh},(see also \cite{Nekrasov:2016gud}
for a  related discussion).  We implicitly assume that such a flux is turned on,  if necessary.

In each case we can discard the $D$ term condition $\mu_{\mathbb{R}}(\zeta) =0$
 (or the real component of the hyper-K\"ahler moment map ) with $\zeta >0$ in favor of the following stability condition;
$$
\hbox {If a subspace $K' \subset K$ satisfies $I(N) \subset K'$ and $B_a(K') \subset K'$, then $K'=K$,}
$$
with the gauge symmetry being complexified to $GL(k, \mathbb{C})$. 
We can show that the $F$-term condition $\mu_{\mathbb{C}}=0$  implies that $B_a$ are commuting $\left[ B_a,  B_b \right] =0$.
In the case of $d=4$ it follows from $\mathrm{Tr}~(\mu_{\mathbb{C}})^2 =0$. In other cases we use
the stability condition to show the vanishing of $J$ or $Y$. Then the stability condition implies that
the vector space $K$ is spanned by action of $B_a$ on the subspace (\lq vacuum\rq) $I(N)$;
\beq\label{cyclic} 
K = \mathbb{C} [ B_a ] \cdot I(N). 
\eeq
We will use this property, when we compute the equivariant character of the tangent space in section 5.

The formal complex dimensions of the moduli space are
computed by subtracting the gauge degrees of freedom and constraints from
the total number of components of matrices;
\begin{enumerate}
\item $d=2$
$$
2 k^2 + 2nk - k^2 - k^2 = 2nk,
$$
\item $d=3$
$$
3 k^2 + k^2 + nk - k^2 - 3 k^3 - nk = 0,
$$
\item $d=4$
$$
4 k^2 + nk - k^2 - 3 k^3 = nk.
$$
\end{enumerate}
Note that if we did not introduce $Y$ in $d=3$, the computation was
$$
3 k^2 + nk - k^2 - 3 k^3 = (n-k) k
$$
and we cannot have a good moduli space. 
Since the dimensions are not necessarily even for $d=4$, the moduli space cannot be
hyperK\"ahler. In fact, for $d>2$ the moduli space is not smooth and the tangent space
only has a virtual meaning.

When $n=1$ which corresponds to abelian gauge theory on $Dp$-branes,
we expect the moduli space is mathematically equivalent to the Hilbert scheme
of $k$ points on $\mathbb{C}^d$. It is known when $d>2$ it is qualitatively
different from the case of $\mathbb{C}^2$ \cite{Okounkov:2015spn}. It is desirable to
clarify the meaning of the generalized ADHM conditions from the viewpoint of 
the Hilbert scheme of $k$ points on $\mathbb{C}^d$.


\section{Matrix model description}

One can construct a cohomological matrix model by imposing ADHM type BPS conditions 
as gauge fixing condition of cohomological matrix model, which is achieved in BRST manner. 
In the case of $d=2$ ADHM constraints are obtained as hyperK\"ahler moment maps and 
this leads to integration over the Higgs branch of supersymmetric quantum mechanics \cite{Moore:1997dj}, 
\cite{Moore:1998et}. Equivariant localization of topological (BRST) symmetry allows us 
to compute the partition function as a residue integral over the eigenvalues (diagonal elements) of the matrix . 
It turns out that the poles of the residue integral are labeled by
partitions and after the reside integral, we obtain a summation over the partitions.

\subsection{$d=2$ (From localization to Macdonald polynomials)}

Let $\{ x_i \}_{i=1}^k$ be the Cartan variables of $GL(k)$ or the eigenvalues of $k$ by $k$ matrices. 
The equivariant integration over the instanton moduli space $\mathcal{M}_{n,k}$ is reduced to 
a contour integral 
\beq
Z_k = \frac{1}{k!} \oint \prod_{i=1}^k \frac{dx_i}{2\pi \sqrt{-1} x_i} z_k(x_i, u_\alpha, q_1, q_2),
\eeq
where we have divided the integral by the order of the Weyl group (we will order the eigenvalues) and
$\{ u_\alpha = e^{a_\alpha} \}_{\alpha=1}^n$ is the Cartan variables for $GL(n)$ symmetry coming from $n$ $D4$ branes. 
The parameters $q_i = e^{\epsilon_i}$ are $\Omega$-background parameters or the equivariant parameters of
the torus action $(z_1, z_2) \to (q_1 z_1, q_2 z_2)$ on $\mathbb{C}^2$. 
The full partition function is 
\beq
Z^{\mathrm{4D}} (u_\alpha, q_i ; \gq) = 1 + \sum_{k=1}^\infty \gq^k Z_k
\eeq
and we will see by introducing the power sum function $p_n(x)$ of the eigenvalues,
the integrand $z_k(x_i, u_\alpha, q_1, q_2)$ allows a plethystic expression.
Note that we may identify $\log z_k(x_i, u_\alpha, q_1, q_2)$ as an effective action of the matrix model. 
The contributions to $z_k(x_i, u_\alpha, t_1, t_2)$ are evaluated as follows\footnote{These 
contributions are in one to one correspondence with the terms in the equivariant 
character to be given in the next section.};

\begin{itemize}
\item
Jacobian (Vandelmonde determinant) from the change of variables to diagonal variables;
\beq
\prod_{i \neq j} \left( 1 - \frac{x_i}{x_j} \right).
\eeq
This factor is also regarded as the contribution of $GL(k)$ gauge symmetry of ADHM constraints. 
\item
Contribution of ADHM constraints;
\beq
\prod_{i,j}  \left( 1 - q_1 q_2 \frac{x_i}{x_j} \right) = (1- q_1 q_2)^k \prod_{i \neq j}  \left( 1 - q_1 q_2 \frac{x_i}{x_j} \right).
\eeq
\item
Contribution of matrix variables $B_{1,2}, I, J$;
\beq
\prod_{i,j}  \left( 1 - q_a \frac{x_i}{x_j} \right)^{-1} = (1- q_a)^{-k } \prod_{i \neq j}  \left( 1 - q_a \frac{x_i}{x_j} \right),
\eeq
from $B_a$ and 
\beq
\prod_{i=1}^k \prod_{\alpha=1}^n \left( 1 - \frac{x_i}{u_\alpha} \right)^{-1},  
\qquad
\prod_{i=1}^k \prod_{\alpha=1}^n \left( 1 - q_1 q_2 \frac{u_\alpha}{x_i} \right)^{-1},
\eeq
from $I$ and $J$.
\end{itemize}
Let us rescale the variable $u_\alpha \to \sqrt{q_1 q_2} u_\alpha$ to make the last two contributions
symmetric;
\beq
\prod_{i=1}^k \prod_{\alpha=1}^n \left( 1 - \sqrt{q_1 q_2} \frac{x_i}{u_\alpha} \right)^{-1},  
\qquad
\prod_{i=1}^k \prod_{\alpha=1}^n \left( 1 - \sqrt{q_1 q_2} \frac{u_\alpha}{x_i} \right)^{-1}.
\eeq

In terms of the function 
\beq
S(z) := \frac{(1-z) (1 - q_1 q_2 z)}{(1- q_1 z)(1 - q_2 z)},
\eeq
we can write the integrand as follows;
\beq
z_k(x_i, a_\alpha, q_1, q_2) = \left( \frac{1- q_1 q_2}{(1-q_1)(1-q_2)}\right)^k 
\frac{\displaystyle{\prod_{i \neq j}} S \left( \frac{x_i}{x_j }\right)}
{\displaystyle{\prod_{i=1}^k \prod_{\alpha=1}^n }\left( 1 - \sqrt{q_1 q_2} \frac{x_i}{u_\alpha} \right)
\left( 1 - \sqrt{q_1 q_2} \frac{u_\alpha}{x_i} \right)}.
\eeq

Now in terms of the power sum variables $p_m = \displaystyle{\sum_{i=1}^k} x_i^m$, we can rewrite
the measure $z_k(x_i, u_\alpha, q_1, q_2)$ for the contour integral in a plethystic form;
\beqa
\log ( z_k(x_i, u_\alpha, q_1, q_2) )
&=& \sum_{m=1}^\infty \frac{1}{m} (q_1^m + q_2^m - q_1^m q_2^m)  \sum_{i,j =1}^k  \left( \frac{x_i}{x_j} \right)^m
-  \sum_{m=1}^\infty \frac{1}{m} \sum_{i \neq j} \left( \frac{x_i}{x_j} \right)^m \CR
&& + \sum_{m=1}^\infty \frac{(\sqrt{q_1 q_2})^m}{m} \sum_{i=1}^k \sum_{\alpha=1}^n \left[ \left(\frac{x_i}{u_\alpha}\right)^m  
 + \left(\frac{u_\alpha}{x_i}\right)^m  \right] \CR
 &=& k \sum_{m=1}^{\infty} \frac{1}{m} - \sum_{m=1}^\infty \frac{1}{m} (1- q_1^m)(1 - q_2^m) p_m p_{-m} \CR
&& + \sum_{m=1}^\infty \frac{(\sqrt{q_1 q_2})^m}{m} \sum_{\alpha=1}^n ( p_m u_\alpha^{-m} +  p_{-m} u_\alpha^{m}).
\eeqa

Using the holonomy variables
\beq
U_m := \sum_{\alpha=1}^n u_\alpha^m
\eeq
of $U(n)$ gauge fields, we have
\beqa
&&\log ( z_k(x_i, u, q_1, q_2) ) = \log \Lambda^k \CR
&&~+\sum_{m=1}^{\infty} \frac{1}{m} \Big[ - (1- q_1^m)(1 - q_2^m) p_m p_{-m} 
+ {(\sqrt{q_1 q_2})^m} ( p_m U_{-m} +  p_{-m} U_m) \Big],
\eeqa
where we have introduced $\log \Lambda := \displaystyle{\sum_{m=1}^\infty} \frac{1}{m}$.
By the change of variables 
\beq \label{change}
\alpha_m := \frac{(\sqrt{q_1q_2})^m}{1- q_1^m} U_m - (1- q_2^m) p_m,
\eeq
we can eliminate linear terms in $p_m$ to obtain
\beq
\log ( z_k(x_i, u, q_1, q_2) ) =  \log \Lambda^k + 
\sum_{m=1}^{\infty} \frac{1}{m} \left[- \frac{(1- q_1^m)}{(1 - q_2^{-m})} \alpha_m \alpha_{-m} 
+ \frac{1}{( 1 - q_1^{-m}) ( 1 - q_2^{-m})} \right].
\eeq
Thus we have
\beq\label{macdonald}
Z_k = \frac{1}{k!} \frac{\Lambda^k} {\prod_{i.j =1}^\infty (1- q_1^i q_2^j)} 
\oint \prod_{i=1}^k \frac{dx_i}{2\pi \sqrt{-1} x_i} 
\exp \left( - \sum_{m=1}^\infty \frac{1}{m} \frac{(1- q_1^m)}{(1 - q_2^{-m})} \alpha_m \alpha_{-m} \right).
\eeq
We may eliminate $\Lambda$ by the renormalization of
the instanton expansion parameter $\gq$. The universal factor $\prod_{i.j =1}^\infty (1- q_1^i q_2^j)^{-1}$
shoud be identified with the perturbative factor.

In the abelian case the contour integral \eqref{macdonald} is related to the inner product for Macdonald polynomials
\cite{Carlsson:2013jka}. To see it, we should note that the poles of the contour integral are 
labeled by partitions $\lambda$ with $|\lambda| =k$ and the position of poles are given by
\beq
x_i = u \cdot q_1^{a - \frac{1}{2}} q_2^{b - \frac{1}{2}}, \qquad (a,b) \in \lambda,
\eeq
where $u=U_1$ and we have $U_m = u^m$ for the abelian case. 
Hence the power sum takes the following values at the poles;
\beq
p_1^{(\lambda)} = u \sum_{a=1}^{\ell(\lambda)} 
\sum_{b=1}^{\lambda_a}  q_1^{a - \frac{1}{2}} q_2^{b - \frac{1}{2}},
\eeq
and 
\beq
\alpha_1^{(\lambda)}  = u \sqrt{q_1 q_2} \sum_{i=1}^\infty q_1^{i-1} q_2^{\lambda_i}.
\eeq
Thus we recover the topological locus;
\beq
\xi_i = u q_1^{i-\frac{1}{2}} q_2^{\lambda_i + \frac{1}{2}}.
\eeq
This also explains an implication of the change of variables \eqref{change}.
In summary after the contour integration we have
\beq
Z_k = \frac{1}{k!} \frac{\Lambda^k} {\prod_{i.j =1}^\infty (1- q_1^i q_2^j)} 
\sum_{|\lambda| =k} 
\exp \left( - \sum_{m=1}^\infty \frac{1}{m} \frac{(1- q_1^m)}{(1 - q_2^{-m})} \alpha_m^{(\lambda)} \alpha_{-m}^{(\lambda)} \right).
\eeq
Note that the measure factor coincides with the $(q,t)$-deformed Vandermonde determinant
\beqa
\Delta_{q,t}(\xi)^2 &:=& \exp \left( \sum_{k=1}^\infty \frac{1}{k} \frac{1-q^k}{1- t^{k}} (N - \alpha_k \alpha_{-k})  \right) \CR
&=& \prod_{n=1}^\infty \prod_{1 \leq a \neq b \leq N} \frac{1 - t^n \xi_a/\xi_b}{ 1- qt^n \xi_a/\xi_b}
\eeqa
with $(q,t)=(q_1, q_2^{-1})$.
This is employed to define the inner product on the space of symmetric polynomials 
that leads to Macdonald polynomials \cite{Macdonald}.

The integrand of the residue integral can be expressed in term of the plethystic exponential
and taking the logarithm we may recognize \lq\lq effective\rq\rq\ action for eigenvalues,
which is in turn expressed by the power sum. Then the integral can be related to
the inner product for the Macdonald polynomials, This also means the effective action 
is bilinear in the power sums (the free boson operators). 
To construct refined topological vertex we have to insert a vertex operator.
It is curious that the insertion induces the interaction term in the effective action.

\subsection{$d=3$}

From the ADHM type conditions, we can similarly obtain a contour integral representation of 
the partition function with instanton number $k$;
\beq
Z_k = \frac{1}{k!} \oint \prod_{i=1}^k \frac{dx_i}{2\pi \sqrt{-1} x_i} z_k(x_i, u_\alpha, q_a),
\eeq
where 
\beq
z_k(x_i, u_\alpha, q_a) = \frac{\displaystyle{\prod_{i=1}^k \prod_{\alpha=1}^n} \left( 1 - q_1 q_2 q_3\frac{u_\alpha}{x_i} \right)
\displaystyle{\prod_{i \neq j}} \left( 1 - \frac{x_i}{x_j} \right) 
\displaystyle{\prod_{1\leq a < b \leq 3}\prod_{i,j}}  \left( 1 - q_a q_b \frac{x_i}{x_j} \right)}
{\displaystyle{\prod_{i=1}^k \prod_{\alpha=1}^n} \left( 1 - \frac{x_i}{u_\alpha} \right)
\displaystyle{\prod_{a=1,2,3}\prod_{i,j}}  \left( 1 - q_a \frac{x_i}{x_j} \right)
 \displaystyle{\prod_{i,j}}  \left( 1 - q_1 q_2 q_3 \frac{x_i}{x_j} \right)}.
\eeq
It is curious to see the role of the pole at $x_j = q_1 q_2 q_3 x_i$ in the contour integral.
An analogous computation to the case of $d=2$ leads the following plethystic form of the measure;
\beqa\label{6dmeasure}
&&\log ( z_k(x_i, u, q_a) ) =   \log \Lambda^k \CR
&&~~+ \sum_{m=1}^{\infty} \frac{1}{m} \big[ - (1- q_1^m)(1 - q_2^m)(1 - q_3^m)  p_m p_{-m} 
+ {(\sqrt{q_1 q_2 q_3})^m} ( p_m U_{-m} -  p_{-m} U_m ) \big]. \CR
\eeqa
The crucial change here is the relative sign in the linear terms,
which prevents us to make a complete square by the change of variable like \eqref{change}.
The flip of the relative sign causes an asymmetry in exchanging the positive modes and 
the negative modes. As will be discussed in the next section this seems to be related to the
fact in contract to the case of $d=2$, we do not have hyperK\"ahler (holomorphic symplectic) 
structure any more when $d=3$.

\subsection{$d=4$}

We obtain
\beq
z_k(x_i, u_\alpha, q_a) = \frac{\displaystyle{\prod_{i \neq j}} \left( 1 - \frac{x_i}{x_j} \right) 
\displaystyle{\prod_{1\leq a < b \leq 3}\prod_{i,j}}  \left( 1 - q_a q_b \frac{x_i}{x_j} \right)}
{\displaystyle{\prod_{i=1}^k \prod_{\alpha=1}^n} \left( 1 - \frac{x_i}{u_\alpha} \right)
\displaystyle{\prod_{a=1}^3 \prod_{i,j}}  \left( 1 - q_a \frac{x_i}{x_j} \right) 
\displaystyle{\prod_{i,j}} \left( 1 - q_4^{-1} \frac{x_i}{x_j} \right)}.
\eeq
As we will argue in the next section, we have to choose a \lq\lq chiral-half\rq\rq\ of the full Euler character,
which depends on the ordering of the set $\{ (ab) \vert 1 \leq a \neq b \leq 4\}$.
Here we chose $\{ (12), (13), (23) ; (14), (24), (34) \}$ by taking $z_4$ as a \lq\lq preferred\rq\rq\ direction.
As argued in \cite{Nekrasov:2016gud}, due to the choice of the ordering, we should be careful with the order of 
the contour integral. 

Using the Calabi-Yau condition $q_1 q_2 q_3 q_4 =1$, we can obtain a plethystic form
of $z_k(x_i, u_\alpha, q_a)$ as follows;
\beqa
&&\log ( z_k(x_i, u_\alpha, q_a) ) =   \log \Lambda^k 
+ \sum_{m=1}^{\infty} \frac{1}{m} \big[ q_1^m + q_2^m + q_3^m + q_4^{-m} \big] p_m p_{-m}  \CR
&&~~~ - \sum_{m=1}^{\infty} \frac{1}{m} \big[1+ q_1^m q_2^m + q_1^m q_3^m + q_2^m q_3^m \big] p_m p_{-m} 
+ \sum_{m=1}^{\infty} \frac{1}{m} p_m U_{-m} \CR
&&=  \log \Lambda^k + \sum_{m=1}^{\infty} \frac{1}{m} p_m U_{-m} 
- \sum_{m=1}^{\infty} \frac{1}{m} (1-q_1^m)(1-q_2^m)(1-q_3^m) p_m p_{-m}.
\eeqa
Remarkably this is quite close to \eqref{6dmeasure}. Since this is a \lq\lq chiral-half\rq\rq\ of the full Euler character,
only the negative modes $U_{-m}$ appear.


\section{Equivariant character of (virtual) tangent space at fixed points}

The fixed points of the toric action of $T^d$ on $\mathbb{C}^d$ and the Cartan subalgebra of 
the gauge symmetry $G_{C}$ are labelled by $n$-tuple of $(d-1)$ partitions. 
In terms of the equivariant parameters $u_\alpha := e^{a_\alpha}$ of the Cartan
subgroup of $U(n)_C$, the character of the vector space $N$ (the Chan-Paton
bundle for the background $Dp$ branes) is\footnote{By the abuse of notation 
we use the same notation for the character.}
\beq
N = \sum_{\alpha=1}^n u_\alpha
\eeq
Then from the structure of the vector space $K$ \eqref{cyclic}, its equivariant character at the fixed point
$\{ \pi_\alpha\}_{\alpha=1}^n$ is
\beq
K_{\pi} = \sum_{\alpha=1}^n u_\alpha \left( \sum_{(i,j,k) \in \pi^\alpha } q_1^{1-i} q_2^{1-j} q_3^{1-k} \right),
\eeq
where for illustration we write the formula for $d=3$, but generalization to other cases, where $\pi$ 
stands for partition ($d=2$) and solid partition ($d=4$), should be clear. 
With these basic ingredients we can compute the (Euler) characters of the deformation complex 
for the ADHM type equation in each dimension.

\subsection{$d=2$}

The fixed points are labelled by $n$-tuple of partitions (colored Young diagrams) $\lambda_\alpha$ and 
\beq\label{4Deqch}
\chi_{4D} (u_\alpha, q_i) = N^{*} K + q_1 q_2 K^{*} N - (1-q_1)(1- q_2) K^{*}K,
\eeq
where the positive contributions $N^{*} K$ and $t_1 t_2 K^{*} N$ come from $I$ and $J$, $(q_1 + q_2) K^{*}K$ from $B_{1,2}$,
while the negative ones $-q_1q_2  K^{*}K$ from the $F$-term constraint and $-K^{*}K$ from the gauge symmetry. 
The difference of the numbers of positive coefficients $(+1)$ and negative coefficients $(-1)$ is $2nk$, which is exactly
the (complex) dimensions of the tangent space. After cancellations only positive term survive and when $n=1$
it has a nice combinatorial formula \cite{Nakajima};
\beq\label{combi-ch}
\chi_{4D} (q_i) = \sum_{s \in \lambda} \left( q_1^{-\ell(s)} q_2^{a(s) +1} + q_1^{\ell(s) +1} q_2^{-a(s)} \right),
\eeq 
where $a(s)$ and $\ell(s)$ are the arm and the leg length of the box $s$ in the Young diagram $\lambda$. 
Note that in abelian case the dependence on $u_\alpha$ disappears. 
In the non-abelian $(n>1)$ case the fixed points are labeled by $n$-tuple of Young diagrams $\vec{\lambda} =( \lambda^\alpha )$
and we need the arm and the leg length of the box $s=(i,j) \in \lambda$ with respect to a second Young diagram $\mu$;
\beq
a_\mu(i,j) := \nu_i - j,  \qquad  \ell_\mu(i,j) := \nu_j^{\vee} - i.
\eeq
Then an explicit formula for the equivariant character is
\beqa
\chi_{4D} (u_\alpha, q_i) &=& \sum_{\alpha,\beta=1}^{n} N_{\alpha \beta}, \CR
N_{\alpha \beta}  (u_\alpha, q_i) &=& \frac{u_\beta}{u_\alpha}  \left( \sum_{s \in \lambda^\alpha} 
q_1^{\ell_{\lambda^\beta}(s)} q_2^{a_{\lambda^\alpha}(s)+1}
+\sum_{t \in \lambda^\beta} q_1^{\ell_{\lambda^\alpha}(s)+1}  q_2^{-a_{\lambda^\beta}(s)} \right).
\eeqa

\begin{figure}[h]
\begin{center}
\begin{picture}(50,140)(-40,0)
\unitlength 1.1mm
\thicklines
\put(-28,10){\circle{6}}
\put(32,10){\circle{6}}
\put(-30,30){\line(1,0){4}}
\put(-30,34){\line(1,0){4}}
\put(-30,30){\line(0,1){4}}
\put(-26,30){\line(0,1){4}}
\put(30,30){\line(1,0){4}}
\put(30,34){\line(1,0){4}}
\put(30,30){\line(0,1){4}}
\put(34,30){\line(0,1){4}}
\put(-28,30){\vector(0,-1){10}}
\put(-28,22){\line(0,-1){9}}
\put(31,30){\vector(0,-1){10}}
\put(31,22){\line(0,-1){9}}
\put(33,13){\vector(0,1){10}}
\put(33,22){\line(0,1){8}}
\put(-38,10){\arc{15}{0.2}{6.1}}
\put(22,10){\arc{15}{0.2}{6.1}}
\put(42,10){\arc{15}{0}{2.9}}
\put(42,10){\arc{15}{3.3}{6.5}}
\put(-38,17.5){\vector(1,0){1.5}}
\put(22,17.5){\vector(1,0){1.5}}
\put(42,17.5){\vector(-1,0){1.5}}

\put(27,20){$I$}
\put(35,20){$J$}
\put(8,10){$B_1$}
\put(52,10){$B_2$}
\end{picture}
\end{center}
\caption{ADHM quiver [right] as the double of the Jordan (framed $\widehat{A_0}$) quiver [left]}
\label{fig:ADHM}
\end{figure}
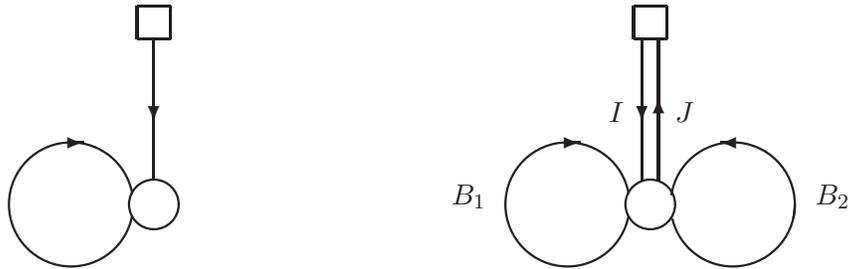


Introducing the following polarization\footnote{By definition a polarization of a symplectic manifold 
$X$ is an equivariant $K$ theory class $P = T^{1/2} X \in K_{T}(X)$, such that the tangent space is 
represented as $TX = P + \hbar P^{*}$.}
\beq\label{por2}
P_2 (u_\alpha, q_i) = N^{*} K + (q_1 -1) K^{*} K,
\eeq
and the notation $\hbar := q_1 q_2$ we can express the character as follows;
\beq
P_2 + \hbar P_2^{*} = N^{*} K - (1-q_1) K^{*} K + q_1q_2 (K^{*} N - (1- q_1^{-1})  K^{*} K) = \chi_{4D}.
\eeq
Note that $\hbar$ is the scaling factor of the symplectic form $\omega = dz_1 \wedge dz_2$.
This decomposition of the equivariant character $\chi_{4D}$ reflects the fact that 
the moduli space of ADHM matrix model is an example of Nakajima quiver varieties, 
which is defined as a hyperK\"ahler quotient. The relevant quiver is called 
Jordan quiver which consists of a single vertex with a single loop (Fig.\ref{fig:ADHM}).
More precisely it is the framed Jordan quiver with a framing of $\mathbb{C}^n$.
When $n=1$ or $U(1)$ gauge theory, the associated quiver variety is
nothing but the Hilbert scheme $\mathrm{Hilb}_k \mathbb{C}^2$ 
of $k$-points on $\mathbb{C}^2$, where $k$ is physically the number of $D0$-branes
or the instanton number of anti-self-dual connection. From this viewpoint 
the moduli space has the structure of a cotangent bundle and 
the polarization $P_2$ represents the contribution of the base space described 
by the Jordan quiver, where we subtract $K^{*} K$ coming from the gauge symmetry.
Then the second term corresponds to the fiber of the cotangent bundle and 
the multiplication of the weight $\hbar$ is necessary.

The equivariant character \eqref{4Deqch} is also derived from the equivariant Chern character 
of the universal bundle $\mathcal{E}$ \cite{Atiyah:1984tf} \cite{Kanno:1988wm}.
The virtue of this derivation is that it is applicable for more general gauge groups of type $SO$ and $Sp$
\cite{Shadchin:2004yx}. To construct the universal bundle $\mathcal{E}$, 
let $m^I$ be local coordinates on the moduli space of instantons. 
The tangent space of the moduli space is spanned by solutions to the linearized equations with 
a gauge fixing condition. Let $\{ \psi_\mu^I (x,  m) \}$ denote a basis of the tangent space 
at $m \in \mathcal{M}_{\mathrm{inst}}$.
For a family of instantons $A_\mu(x, m)$ parametrized by $m$, we have
\beq
\frac{\partial A_\mu(x,m)}{\partial m_I} = h_{IJ} \psi_\mu^J + D_\mu \alpha_I.
\eeq
Since the derivative of $A_\mu(x, m)$ does not necessarily satisfy the gauge fixing condition 
we need a compensating gauge transformation $D_\mu \alpha_I$. With an appropriate choice of
the gauge fixing condition, for example $(D^{*})^\mu \psi_\mu^I (x,  m) =0$, we can find a unique $\alpha_I$.
Combining $A_\mu$ with the parameter of the compensating gauge
transformation $\alpha_I$, we can define a one form ${\mathcal A}(x,m) = A_\mu dx^{\mu} + \alpha_I dm^I$
which can be regarded as a connection of the universal  bundle $\mathcal{E}$ 
on $\mathbb{R}^4 \times \mathcal{M}_{\mathrm{inst}}$ whose fiber is the fundamental representation $
\mathbb{C}^n$ of $U(n)$. In the following we fix a complex structure of the space-time $\mathbb{R}^4$ 
and identify $\mathbb{R}^4 \simeq \mathbb{C}^2$. Then the spinor bundle $S^{+} \oplus S^{-}$ on
$\mathbb{R}^4$ is naturally identified with the space of $(0,k)$ forms $\Lambda^{(0,0)} \oplus
\Lambda^{(0,1)} \oplus \Lambda^{(0,2)}$ on $\mathbb{C}^2$~\footnote{In general the spinors 
on a Calabi-Yau manifold are equivalent to $(0,k)$ forms, where the chirality of the spinor 
corresponds to the parity of $k$.}. With this identification the Dirac operator
is translated to $\bar\partial$ operator.

The equivariant Chern character of the universal bundle $\mathcal{E}$ is computed 
as the Euler character of the complex
\beq\label{complex}
0 \longrightarrow K \otimes \Lambda^{(0,0)} \xrightarrow{~~~\tau_z~~~} 
K \otimes \Lambda^{(0,1)} \oplus N \otimes  \Lambda^{(0,2)}
\xrightarrow{~~~\sigma_z~~~} K \otimes \Lambda^{(0,2)} \longrightarrow 0,
\eeq
where
\beq
\tau_z =
\left( \begin{array}{c}
B_1 - z_1 \\ B_2 -z_2  \\ J
\end{array} \right),
\qquad
\sigma_z = \left( \begin{array}{ccc} - (B_2 -z_2)  & B_1-z_1 & I 
\end{array} \right),
\eeq
and the ADHM condition guarantees \eqref{complex} is a complex; $\sigma_z \circ \tau_z =0$.
One can also check $\mathrm{Ker}~\sigma_z = \mathrm{Coker}~\tau_z =0$ \cite{Nakajima}. 
Taking the alternating sum, we obtain
\beq
\mathrm{Ch}_q(\mathcal{E}) (u_\alpha; q_i)  =  N(u_\alpha) - (1-q_1)(1-q_2) K(u_\alpha; q_i).
\eeq
Now the equivariant version of the index theorem tells the equivariant index of the Dirac operator coupled 
with the adjoint bundle $\mathcal{E} \otimes \mathcal{E}^{*}$ is\footnote{
The Dirac operator on complex manifold is related to the the $\bar\partial$ operator
by the twist of the squrare root of the determinant of the tangent bundle, which is trivial
for Calabi-Yau case.}
\beq
\mathrm{Ind}_q~\bar\partial_{\mathcal{E} \otimes \mathcal{E}^{*}}
= \int_{\mathbb{C}^2} \mathrm{Ch}_q(\mathcal{E} \otimes \mathcal{E}^{*})
\mathrm{Td}_q (\mathbb{C}^2),
\eeq
where the equivariant version of the Todd class is
\beq
\mathrm{Td}_q (\mathbb{C}^2) = \frac{x_1 x_2}{(1- e^{x_1})(1- e^{x_2})},
\eeq
where
\beq
x_i = \epsilon_i + \delta(z_i)\frac{dz_i \wedge d\bar{z_i}}{2\pi \sqrt{-1}}  
\eeq
are the equivariant Chern roots of the tangent bundle to $\mathbb{R}^4 \simeq \mathbb{C}^2$,
given by equivariantly closed two forms. 
We should use the Chern class of $\mathcal{E} \otimes \mathcal{E}^{*}$, because we consider
the adjoint bundle whose fibre is the adjoint representation of $U(n)$. It shoud be easy to
generalize the computation to the bi-fundamental representation. 
The integration over the space-time $\mathbb{C}^2 \simeq \mathbb{R}^4$ corresponds to
the push-forward for the projection $\pi : \mathbb{R}^4 \times \mathcal{M}_{\mathrm{inst}}
\longrightarrow \mathcal{M}_{\mathrm{inst}}$ and can be evaluated by the localization by
the torus action $(z_1, z_2) \to (q_1z_1, q_2 z_2)$, whose unique fixed point is the origin $z_1=z_2=0$. 
The Hamiltonian of the torus action is $\epsilon_1 |z_1|^2 + \epsilon_2 |z_2|^2$ and
the localization theorem for the equivariant closed forms gives
\beq
\int_{\mathbb{C}^2} \mathrm{Ch}_q(\mathcal{E} \otimes \mathcal{E}^{*})
\mathrm{Td}_q (\mathbb{C}^2)
= \frac{\mathrm{Ch}_q(\mathcal{E} \otimes \mathcal{E})
\mathrm{Td}_q (\mathbb{C}^2) \vert_{(0,0)}} {\epsilon_1 \epsilon_2}
=  - \frac{N^* N} {(1-q_1)(1-q_2)} + \chi_{4D},
\eeq
where the first term, which survives even $k=0$, is regarded as a perturbative part.

\subsection{$d=3$}

The fixed points are labelled by $n$-tuple of plane partitions and 
\beq\label{6Deqch}
\chi_{6D} (u_\alpha, q_i) = N^{*} K - q_1 q_2 q_3 K^{*} N - (1-q_1)(1- q_2)(1- q_3) K^{*}K,
\eeq
where $N^{*} K, (q_1 + q_2 + q_3) K^{*}K$ and $q_1q_2 q_3 K^{*}K$ come from dynamical matrix variables
$I, B_{1,2,3}$ and $Y$, while $-(q_1q_2+q_2 q_3 + q_3 q_1) K^{*}K$ from the $F$-term constraints 
and $-K^{*}K$ from the gauge symmetry. 
Finally $- q_1 q_2 q_3 K^{*} N$ comes from the constraint $\mu_B =0$. 
When we impose the Calabi-Yau condition $q_1 q_2 q_3 =1$, the character is anti-self dual $\chi_{6D} + \chi_{6D}^{*} =0$,
which is a consequence of the Serre duality. By the anti-self duality, the measure on the space of plane partitions 
becomes uniform (up to sign), in fact it is $(-1)^{nk}$. Hence the partition function reduces to the MacMahon function. 

Now the analogue of the polarization \eqref{por2} in $d=3$ is
\beq
P_3 (u_\alpha, q_i):= N^{*} K + (q_1 + q_2 + q_3 -1)  K^{*} K,
\eeq
and we set $\hbar = q_1 q_2 q_3$, then we have
\beq\label{decomp3}
\chi_{6d} = P_3 - \hbar P_3^{*}.
\eeq
Note that the relative sign between $P$ and $P^{*}$ should be negative for odd $d$.
Consequently the interpretation of the decomposition \eqref{decomp3} is rather different from the case $d=2$.
Namely \eqref{decomp3} reflects what is called symmetric obstruction theory in mathematics,
where the first term corresponds to the deformation space of matrix variables coming from the framed quiver
with a single vertex and three loops with the subtraction of gauge symmetry, 
while the second term is the contributions from the obstruction space, which are represented by
anti-ghosts for constraints and the secondary ghost for the gauge symmetry. 
The symmetric obstruction theory gives a moduli space of virtual dimension zero.

\subsection{$d=4$}

The fixed points are labelled by $n$-tuple of solid partitions and 
\beq\label{8Deqch}
\chi_{8D} (u_\alpha, q_i) = N^{*} K - (1-q_1)(1- q_2)(1- q_3)K^{*}K = P_4,
\eeq
where $N^{*} K$ and $(q_1 + q_2 + q_3 + q_1q_2q_3) K^{*}K$ come from $I$ and $B_{1,2,3,4}$, 
while $-(q_1q_2+q_2 q_3 + q_3 q_1)   K^{*}K$ from the $F$-term constraints and $-K^{*}K$ from the gauge symmetry. 
Note that  what we called the polarization in lower dimensional cases is obtained as the character of the deformation
complex of ADHM type condition. Namely \eqref{8Deqch} is a \lq\lq chiral-half\rq\rq\ of the full Euler character 
$\chi_E(\mathcal{E}) =  \sum_{i=0}^4 (-1)^i \mathrm{Ext}^{i} (\mathcal{E}, \mathcal{E})$;
$$
\chi_{E} (u_\alpha, q_i) = N^{*} K + K^{*} N  - (1-q_1)(1- q_2)(1- q_3)(1-q_4) K^{*}K = \chi_{8D} + \chi_{8D}^{*},
$$
where we have used $\hbar = q_1 q_2 q_3 q_4=1$. Contrary to the odd dimensional case, the full character is
self dual $\chi_E^{*} = \chi_E$. To define a \lq\lq chiral-half\rq\rq\ of the full Euler character, we use
the real structure of $\mathrm{Ext}^{2} (\mathcal{E}, \mathcal{F})$, which is allowed 
by the Serre duality of $\mathrm{Ext}^{i} (\mathcal{E}, \mathcal{E})$.
This seems consistent with the idea that $d=4$ theory is a holomorphic version of the Donaldson theory \cite{Baulieu:1997jx}. 
By taking a \lq\lq chiral-half\rq\rq\ of the full Euler character we consider a square root of the tangent space
and hence there is an ambiguity of the choice of sign. We can fix it locally,  but the global consistency is a non-trivial issue. 
Mathematically this is the problem of the orientability of the moduli space.


\section{Topological partition function}

As we have emphasized in introduction, all the partition function in the following can be expressed 
as plethystic exponentials.

\subsection{$d=2$ (Abelian $\mathcal{N}=2^{*}$ theory)}

Using the formula \eqref{combi-ch} the partition function of $U(1)$ theory with adjoint matter is 
\beq \label{2dmeasure}
Z_{U(1), \mathrm{adj}}^{4D} (q_a, \mu; \gq)
= \sum_{\lambda} \gq^{|\lambda|} \prod_{s \in \lambda} 
\frac{1- \mu q_1^{-\ell(s)} q_2^{a(s)+1}}{1- q_1^{-\ell(s)} q_2^{a(s)+1}}
\frac{1- \mu q_1^{\ell(s)+1} q_2^{-a(s)}}{1- q_1^{\ell(s)+1} q_2^{-a(s)}},
\eeq
where the parameter $\mu := e^{-m}$ is the equivariant (mass) parameter for the $U(1)$ flavor symmetry of the adjoint matter.
Thus, physically $Z_{U(1), \mathrm{adj}}^{4D}$ is the Nekrasov partition function of $\mathcal{N}=2^{*}$ theory. 
We can show that it has the following plethystic form 
\cite{Iqbal:2008ra},\cite{Poghossian:2008ge},\cite{Awata:2009yc},\cite{Carlsson:2013jka};
\beqa\label{adjoint}  
Z_{U(1), \mathrm{adj}}^{4D} (q_a, \mu; \gq)&=& \hbox{P. E.} \left[ F(q_a, \mu, \gq) \right], \CR
F(q_a, \mu; \gq) &:=& \frac{- \sqrt{\mu \gq}[\mu q_1][\mu q_2]}{ [q_1] [q_2] [\mu\gq] }
= \frac{\gq}{1 - \mu \gq} \frac{(1 - \mu q_1)(1- \mu q_2)}{(1-q_1)(1-q_2)}. 
\eeqa
There are two natural limits for $\mu$; the decoupling limit $\mu \to 0$ and the massless ($\mathcal{N}=4$) limit
$\mu \to 1$. In the latter case the measure on the space of partitions is uniform and we obtain 
the generating function of the counting of partitions;
\beq
Z_{U(1),\mathcal{N}=4}^{4D} (\gq) =  \hbox{P. E.} \left[ \frac{\gq}{1-\gq} \right] = \frac{1}{(\gq;\gq)_\infty}.
\eeq
On the other hand, in the former limit the measure becomes the (refined) Plancherel measure and 
corresponds to the pure $U(1)$ theory, which is geometrically engineered by the conifold geometry;
\beq
Z_{U(1), \mathrm{adj}}^{4D} (q_a; \gq) = \hbox{P. E.} \left[ \frac{\gq/\sqrt{q_1q_2}}{[q_1] [q_2] } \right].
\eeq
When $q_1 = q_2^{-1} =t$ it gives a generalized McMahon function
\beq
Z_{U(1), \mathrm{adj}}^{4D} (t; \gq) 
=\hbox{P. E.} \left[ \frac{- \gq}{[t]^2} \right].
= \prod_{n=1}^\infty (1 + \gq t^{n})^{-n},
\eeq
where $\gq$ plays the role of the K\"ahler parameter of the conifold. 
Thus this example gives a kind of interpolation between the counting of partitions and 
plane partitions.

The formula \eqref{adjoint} can be deduced as follows\footnote{Strictly speaking, we assume that the partition function
has a plethystic form.};
First we note the \lq\lq removable\rq\rq\ boxes of a non-empty partition have vanishing leg and arm length $\ell(s)=a(s)=0$,
because if they have non-empty leg or arm, we cannot remove them from the diagram. From the measure factor in 
\eqref{2dmeasure}, we see the measure on non-empty partition has zeros at $\mu q_1 = 1$ and $\mu q_2 = 1$.
Note that these zeros are preserved under $(q_1, q_2, \mu) \to (q_1^k, q_2^k, \mu^k)$. Thus we conclude that
$F$ has the factor $[\mu q_1] \cdot [\mu q_2]$. Moreover, when $\mu =1$ the measure is independent of $q_1$ and $q_2$.
Hence we arrive at 
\beq
F(q_a, \mu ;\gq) \sim \frac{[\mu q_1] \cdot [\mu q_2]}{[q_1] \cdot [q_2]}.
\eeq
Now let us specialize $q_2 = q_1^{-1}$ and take the limit $q_1 \to 0$. Then
\beqa
Z_{U(1), \mathrm{adj}}^{4D} (q_a, \mu; \gq)
= \sum_{\lambda} \gq^{|\lambda|} \prod_{s \in \lambda} 
\frac{q_1^{h(s)} - \mu}{q_1^{h(s)} - 1}
\frac{1- \mu q_1^{h(s)}}{1- q_1^{h(s)}} 
\to \sum_{\lambda} (\mu \gq)^{|\lambda|},
\eeqa
where $h(s) = \ell(s) + a(s) +1$ is the hook length.
On the other hand
\beq
\frac{[\mu q_1] \cdot [\mu q_2]}{[q_1] \cdot [q_2]} = \frac{[\mu q_1] \cdot [\mu^{-1} q_1]}{[q_1]^2 } \to 1
\eeq
in this limit. Hence we find 
\beq
F(q_a, \mu; \gq) \sim \frac{\mu \gq}{ 1 - \mu \gq} \frac{[\mu q_1] \cdot [\mu q_2]}{[q_1] \cdot [q_2]}
= \frac{- \sqrt{\mu \gq}[\mu q_1][\mu q_2]}{[\mu\gq]  [q_1] [q_2] }.
\eeq

We can also prove \eqref{adjoint} by assuming the invariance of the topological string amplitudes 
under the change of the preferred direction of the refined topological vertex  \cite{Awata:2009yc}. 
When we deduce the refined topological vertex from the equivariant vertex the preferred direction 
is related to the ways of the limit $|q_i | \to \infty$ keeping 
the Calabi-Yau combination of $q_i$ constant (see the next subsection). 
This reminds us of the fact the perturbative string theory can be obtained 
by taking appropriate limits of $M$ theory.

The $\mathcal{N}=2^{*}$ theory can be regarded as $\widehat{A_0}$ quiver gauge theory. 
As we have seen in the last section the quiver for the ADHM equation is the double of the framed 
$\widehat{A_0}$ quiver, and when the framing is $U(1)$ the Nakajima variety is nothing but the Hilbert
scheme of point on $\mathbb{C}^2$. This coincidence seems to be the origin of symmetric property of
the topological partition function derived above.

\subsection{$d=3$}

For abelian case $n=1$ by the localization theorem the partition function is given by the summation over the plane partitions\footnote{
In six dimensional case the natural counting parameter is $(-1)^n \gq$, because with this choice the partition function of $U(n)$ theory
reduces to the $n$-th power of the MacMahon function \cite{Cirafici:2008sn},\cite{Awata:2009dd}
in the Calabi-Yau limit $\hbar \to 1$};
\beq
Z_{U(1)}^{6D} (q_a; \gq) = \sum_{\pi} (-\gq)^{|\pi|} \hat{\bf a} (\chi_{\pi}) = \left\langle \widehat{\hbox{P.E.}} [\chi_{6D}] \right\rangle 
\eeq
where $ \hat{\bf a}$ is defined by
\beq
 \hat{\bf a} ( \sum_i m_i w_i ) = \prod_i [w_i]^{m_i}, \qquad m_i \in \mathbb{Z}, w_i \in T^{\vee}.
\eeq
$m_i$ is the multiplicity of the character (weight) $w_i$ of the torus $T$ that acts on the moduli space. 
Note that $\hat{\bf a}$ gives the character of symmetrized symmetric products;
\beq
\hat{\bf a} ( - \hbox{character of $V$}) = \hbox{character of $\hat{S}^{\bullet} V$},
\eeq
where $\hat{S}^{\bullet} V = (\det V)^{\frac{1}{2}} \cdot S^{\bullet} V$.

It turns out that the partition function allows a plethystic expression  \cite{Nekrasov:JJM}, \cite{Okounkov:2015spn};
\beq\label{DT3U1}
Z_{U(1)}^{6D} (q_a; \gq) =  \hbox{P. E.} \left[ F_1(q_a, \gq) \right],
\eeq
where
\beq\
F_1(q_a;\gq) = \frac{[q_1 q_2] [q_2 q_3] [q_3 q_1]} {[q_1] [q_2] [q_3] [\sqrt{\hbar} \gq] [\sqrt{\hbar}/ \gq]},
\eeq
with $\hbar := q_1 q_2 q_3$. 
It is tempting to identify the parameter $\hbar$ with the mass parameter $\mu$ in the previous example.
In fact both are related to the weight of the line bundle over $X = \mathbb{C}^2$ and  $X = \mathbb{C}^3$, where
the total space is six and ten dimensions, respectively. However, an important difference here is 
the decoupling limit is not well defined, while the Calabi-Yau limit $\hbar \to 1$ is still well-defined. 
It seems this is related to the fact that the tangent space is smooth in $d=2$ case, 
but it is singular (the tangent space only has a meaning as virtual bundle) for $d>2$. 
In the Calabi-Yau limit the partition  function reduces to the MacMahon function;
\beq
F(q_a; \gq) = \frac{[q_1^{-1}] [q_2^{-1}] [q_3^{-1}]} {[q_1] [q_2] [q_3] [\gq] [\gq^{-1}]} = \frac{1}{[\gq]^2}.
\eeq

Another interesting limit is the \lq\lq refined topological vertex limit\rq\rq, where we take
$q_1, q_3 \to 0$ with $|q_1| << |q_3|$ and $q_2 \to \infty$ keeping $\hbar$ constant.
In such a limit $q_3$ corresponds to a preferred direction of the refined topological vertex. 
From the relation
\beq
\frac{[\hbar t]}{[t]} \to 
\begin{cases}
\hbar^{\frac{1}{2}} \qquad t \to \infty \\
\hbar^{-\frac{1}{2}} \qquad t \to  0\\
\end{cases}
\eeq
we find
\beq
F(q_a; \gq) = \frac{-[q_1\hbar] [q_2 \hbar] [q_3 \hbar]} {[q_1] [q_2] [q_3] [\sqrt{\hbar} \gq] [\sqrt{\hbar}/ \gq]}
\to \frac{- \hbar^{-\frac{1}{2}}} { [\sqrt{\hbar} \gq] [\sqrt{\hbar}/ \gq]}
= \frac{-1/\sqrt{q_4 q_5}} { [q_4][q_5]}.
\eeq
which can be identified with the refined conifold amplitude with the K\"aher parameter $-1$.

When $n>1$ (non-abelian case)  the fixed points are labeled by $n$-tuple of plane partitions (colored partitions) 
$\vec{\pi} = (\pi^\alpha)_{\alpha=1}^n$ and the topological partition function is 
\beq\label{DT3Un}
Z_{U(n)}^{6D} (u_\alpha, q_a; \gq) = \sum_{\vec{\pi}} ((-1)^n\gq)^{|\vec{\pi}|} 
\prod_{\alpha,\beta =1}^n \hat{\bf a} (V_{\alpha \beta}),
\eeq
where
\beqa
V_{\alpha \beta} (u_\alpha, q_a)&=&\frac{u_\alpha}{u_\beta} \left( \sum_{(i,j,k) \in \pi^\beta} q_1^{1-i} q_2^{1-j} q_3^{1-k}
- \sum_{(r,s,t) \in \pi^\alpha} q_1^{r} q_2^{s} q_3^{t} \right. \CR
&&~~~ \left. - (1-q_1)(1-q_2)(1-q_3)
 \sum_{(r,s,t) \in \pi^\alpha \atop (i,j,k) \in \pi^\beta } q_1^{r-i} q_2^{s-j} q_3^{t-k} 
\right).
\eeqa
One of the significant properties of $Z_{U(n)}^{6D} (u_\alpha, q_a; \gq)$ so defined is that it is
completely independent of the equivariant parameters $u^\alpha$ for the framing torus,
or the Coulomb branch moduli which physically means the distance of $D6$-branes. 
For lower instanton numbers this crucial property in proved in \cite{Awata:2009dd} by checking 
the vanishing of residues at the possible poles of $Z_{U(n)}^{6D} (q_a; \gq)$
as a rational function in the equivariant parameters $u^\alpha$.
Quite recently it is proved for arbitrary $k$ by examining the contour integral 
representation of the partition function discussed in the last section \cite{Fasola:2020hqa}.
Once we know $Z_{U(n)}^{6D} (q_a; \gq)$ is independent of $u^\alpha$,
we can evaluate the partition function in a judicious scaling of $u^\alpha$,
for example $u^\alpha = L^{\alpha}$ with $L \to \infty$ \cite{Awata:2009dd}.
Then we can see for $\alpha < \beta$ \cite{Fasola:2020hqa};
\beq
\lim_{L \to \infty} \hat{\bf a} (V_{\alpha \beta}) \hat{\bf a} 
(V_{\beta \alpha}) \vert_{u_\alpha = L^\alpha}
= (- \hbar^{\frac{1}{2}})^{|\pi^\beta| - |\pi^\alpha|},
\eeq
which implies
\beqa\label{factorization}
\lim_{L \to \infty} Z_{U(n)}^{6D} (L^\alpha, q_a; \gq) 
&=& \sum_{\vec{\pi}} ((-1)^n \gq)^{|\vec{\pi}|} 
\prod_{\alpha =1}^n \hat{\bf a} (V_{\alpha \alpha})
\prod_{1 \leq \alpha < \beta \leq n}(- \hbar^{\frac{1}{2}})^{|\pi^\beta| - |\pi^\alpha|} \CR
&=& \sum_{\vec{\pi}} \prod_{\alpha =1}^n ((-1)^n \gq)^{|\pi^\alpha|} 
 \hat{\bf a} (V_{\alpha \alpha})(- \hbar^{\frac{1}{2}})^{(-n-1 + 2\alpha)|\pi^\alpha|} \CR
&=& \prod_{\alpha =1}^n Z_{U(1)}^{6D} (q_a; \gq \hbar^{\alpha - \frac{n+1}{2}}).
\eeqa
Hence we have a factorization of $U(n)$ partition function \eqref{DT3Un} into
a product of $n$ $U(1)$ partition functions with shifted instanton number counting parameters 
\cite{Benini:2018hjy},\cite{Nekrasov:2018xsb}.
This factorization is surely relies on the independence of $Z_{U(n)}^{6D} (q_a; \gq)$ of the Coulomb moduli
and is related to the orbifold action of $\mathbb{Z}_n$ on the transverse direction to $D6$-branes.

In fact we can use the following identity for generic variables $z_1, z_2$ 
to derive the $U(n)$ partition function \eqref{DT3U1} from \eqref{factorization};
\beq\label{Zn}
\sum_{\ell=1}^n \frac{1}{[z_1^{n+1-\ell} z_2^{1-\ell}] [z_1^{\ell-n} z_2^\ell]}
=
\frac{1}{n} \sum_{\ell=0}^{n-1} \frac{1}{[\omega^\ell z_1][\omega^{-\ell} z_2]} 
=
\frac{[(z_1z_2)^n]}{[z_1 z_2][z_1^n] [z_2^n]},
\eeq
where $\omega$ is an $n$-th root of unity $\omega^n=1$. 
The second equality of \eqref{Zn} is just a simple consequence of 
\beq
\frac{1}{n} \sum_{\ell=0}^{n-1} \omega^{k\ell} = \delta_{0, k (\mathrm{mod}~n)}.
\eeq
But, as is pointed out in \cite{Nekrasov:2018xsb}, 
it is amusing to note that the first equality of \eqref{Zn} allows a geometric interpretation, 
though it can be also checked by induction on $n$. To see the geometric meaning
let us consider the ALE resolution $\widetilde{S_n} \longrightarrow \mathbb{C}^2/\mathbb{Z}_n$
of the $\mathbb{Z}_n$-orbifold of $\mathbb{C}^2$ where $\mathbb{Z}_n$ acts by
$(z_1, z_2) \to (\omega z_1, \omega^{-1} z_2)$. 
We can compare the equivariant index of the Dirac operator before and after the resolution. 
Recall that the equivariant index of the Dirac operator on $\mathbb{C}^2$ is simply
\beq
\mathrm{Ind}_q D = \frac{1}{[z_1][z_2]}.
\eeq
Then we can recognize the middle of \eqref{Zn} as the orbifold version of the equivariant index. 
On the other hand on the ALE space $\widetilde{S_n}$ there appear 
$n$ fixed points over the origin which is the original fixed point of the torus action. 
For example, when $n=2$, the ALE space $\widetilde{S_2}$ is nothing but the Eguchi-Hanson space
\cite{Eguchi:1978xp}, which is isomorphic to the cotangent bundle of $\mathbb{CP}^1$.
Thus we find two fixed points; the north and the south poles of $\mathbb{CP}^1$.
The weights at these fixed points are exactly those appear on the left hand side of \eqref{Zn}.
Hence it is the blow up version of the index. Mathematically the equality of two versions follows from
the fact the the fibre of the resolution is compact \cite{Nekrasov:2018xsb}.

Since we already know $U(1)$ partition function has a plethystic form \eqref{DT3U1}, 
we can compute a plethystic form of $U(n)$ partition function \eqref{DT3U1} as follows
\beq
Z_{U(n)}^{6D} (q_a; \gq) =  \hbox{P. E.} \left[ F_n(q_a, \gq) \right],
\eeq
where
\beqa
F_n(q_a ; \gq) &:=& \sum_{\alpha=1}^n F_1(q_a ; \gq \hbar^{\alpha - \frac{n+1}{2}}) \CR
&=& \frac{[q_1 q_2] [q_2 q_3] [q_3 q_1]} {[q_1] [q_2] [q_3]} 
\sum_{\alpha=1}^n \frac{1}{[\gq \hbar^{\alpha -\frac{n}{2}}] 
[\gq^{-1}\hbar^{1-\alpha +\frac{n}{2}} ]}.
\eeqa
By applying the formula \eqref{Zn} with $z_1 = \hbar^{\frac{1}{2}} \gq^{-\frac{1}{n}},  
z_2 = \hbar^{\frac{1}{2}} \gq^{\frac{1}{n}}$, we finally obtain
\beq\label{6Dfree}
F_n(q_a ; \gq) = \frac{[q_1q_2] [q_2 q_3] [q_3 q_1] [\hbar^n]} 
{[q_1] [q_2] [q_3] [\hbar] [\hbar^{\frac{n}{2}}  \gq][ \hbar^{\frac{n}{2}} \gq^{-1} ]}.
\eeq

In \cite{Nekrasov:2014nea} the equivariant ($K$ theory or $M$ theory) vertex is defined by
\beq
V(\lambda, \mu, \nu) =  \sum_{\pi \to (\lambda, \mu, \nu)}  (-\gq)^{|\pi|} \hat{\bf a} (\chi_{\pi}),
\eeq
where the summation is taken for the plane partitions with a fixed asymptotic condition $(\lambda, \mu, \nu)$.
Note that $|\pi|$ and $\chi_{\pi}$ have to be regularized by taking the edge contributions into account. 
Recently it has been show that if one of three partitions $(\lambda, \mu, \nu)$ is empty, $V(\lambda, \mu, \nu) $ has
a plethystic expression \cite{Kononov:2019fni}. It is very interesting to see if this property holds for the full vertex.

\subsection{$d=4$}

The following plethysitic form of the partition function is conjectured in \cite{Nekrasov:2017cih}, \cite{Nekrasov:2018xsb};
$$
Z_{U(n)}^{8D}(q_a, \nu_\alpha, \mu_\alpha ; \gq)= \hbox {P.E.} \left[ F(q_a, \frac{\prod \nu_\alpha}{\prod \mu_\alpha}, -\gq) \right],
$$
where
$$
F(q_a, s; \gq) := \frac{[q_1q_2] [q_2 q_3] [q_3 q_1] [s]} {[q_1] [q_2] [q_3] [q_4] [\sqrt{s} \gq][ \gq / \sqrt{s}]}.
$$
Note that  $[q_4] = - [q_1 q_2 q_3]$ due to the Calabi-Yau condition $q_1 q_2 q_3 q_4 =1$. 
$\nu_\alpha = e^{i a_\alpha}$ is the Coulomb branch parameters (associated with the position of $D8$-branes)
for the gauge symmetry $U(n)_{C}$
and $\mu_\alpha = e^{ - m_\alpha}$ is the mass parameter  (associated with the position of $\overline{D8}$-branes)
for the flavor symmetry $U(n)_{F}$. It is remarkable that the final result only depends 
on the ratio ${\prod \nu_\alpha}/{\prod \mu_\alpha}$, which is comparable to the fact that the partition function is 
independent of the Coulomb moduli $u_\alpha$ in six dimensions.

For $U(1)$ theory we can take $\nu=1$ by choosing the position of a single brane as the origin.
With $\mu = e^{-m}$ we find
$$
F(q_a, \mu; \gq) = \frac{-[q_1q_2] [q_2 q_3] [q_3 q_1] [\mu]} {[q_1] [q_2] [q_3] [q_4] [\sqrt{\mu} \gq][ \gq / \sqrt{\mu}]}
=\frac{[q_1q_2] [q_2 q_3] [q_3 q_1] [\mu]} {[q_1] [q_2] [q_3] [q_4] [\sqrt{\mu} \gq][ \sqrt{\mu}/\gq]}.
$$
It seems that we cannot produce the uniform measure on the space of solid partitions by tuning parameters.
This is consistent with the fact that there is no known plethystic formula for the generating function of 
the counting of solid partitions. For example the massless limit $\mu \to 1$ gives a trivial result $F=0$. 

Let us put $q_a= e^{- R \epsilon_a}$ and $\mu = e^{-R m}$ and take the limit $R \to 0$, then 
\beqa
F(t_a, \mu; \gq)  &\to& \exp \left( \frac{m(\epsilon_1 + \epsilon_2)(\epsilon_2 + \epsilon_3)(\epsilon_3 + \epsilon_1)}
{\epsilon_1 \epsilon_2 \epsilon_3 \epsilon_4} \sum_{n=1}^\infty \frac{1}{n} \frac{\gq^n}{(1- \gq^n)}\right) \CR
&=& M(\gq)^{\frac{m(\epsilon_1 + \epsilon_2)(\epsilon_2 + \epsilon_3)(\epsilon_3 + \epsilon_1)}
{\epsilon_1 \epsilon_2 \epsilon_3 \epsilon_4} }.
\eeqa
It is the MacMahon function that appears in this limit. 
The generating function of the counting of the solid partition never appears.

It is interesting that a reduction to six dimensions is achieved by tuning the mass parameters (the positions of $\overline{D8}$-branes) 
which triggers a tachyon condensation of $D8$-$\overline{D8}$ system to $D6$-branes \cite{Nekrasov:2018xsb}. 
The condition is $\nu_\alpha = q_4 \mu_\alpha$, which gives $s=q_4^n$ and we obtain;
$$
F(q_a ; \gq) := \frac{[q_1q_2] [q_2 q_3] [q_3 q_1] [\hbar^n]} {[q_1] [q_2] [q_3] [\hbar] 
 [\hbar^{\frac{1}{2}}  \gq][ \gq \hbar^{-\frac{1}{2}} ]},
$$
where $\hbar = q_1 q_2 q_3 = q_4^{-1}$. Up to sign this agrees with \eqref{6Dfree}.

\newpage


\section*{Acknowledgments}

We would like to thank the organizers of the conference
¡ÈParticle Physics and Mathematical Physics -- 40 years 
after Eguchi-Hanson solution¡É, which was a good opportunity for sharing the memories of Prof. Eguchi. 
We would like to thank H.~Awata, A.~Mironov, A.~Morozov and Y.~Zenkevich for discussion and collaborations. 
The work is supported in part by Grants-in-Aid for Scientific Research
18K03274 and JSPS Bilateral Joint Projects (JSPS-RFBR collaboration)
``Elliptic algebras, vertex operators and link invariants'' from MEXT, Japan.


\begin{thebibliography}{99}

\bibitem{Atiyah:1978ri}
M.~Atiyah, N.~J.~Hitchin, V.~Drinfeld and Y.~Manin,
``Construction of Instantons,''
Phys.\ Lett.\ A \textbf{65}, 185-187 (1978)

\bibitem{Corrigan:1983sv}
E.~Corrigan and P.~Goddard,
``Construction of Instanton and Monopole Solutions and Reciprocity,''
Annals Phys.\  \textbf{154}, 253 (1984)

\bibitem{Witten:1994tz} 
  E.~Witten,
  ``Sigma models and the ADHM construction of instantons,''
  J.\ Geom.\ Phys.\  {\bf 15}, 215 (1995)
  [hep-th/9410052].

\bibitem{Witten:1995gx} 
  E.~Witten,
  ``Small instantons in string theory,''
  Nucl.\ Phys.\ B {\bf 460}, 541 (1996)
  [hep-th/9511030].

\bibitem{Douglas:1995bn} 
  M.~R.~Douglas,
  ``Branes within branes,''
  NATO Sci.\ Ser.\ C {\bf 520}, 267 (1999)
  [hep-th/9512077].
 
\bibitem{Douglas:1996uz} 
  M.~R.~Douglas,
  ``Gauge fields and D-branes,''
  J.\ Geom.\ Phys.\  {\bf 28}, 255 (1998)
  [hep-th/9604198].

\bibitem{Nekrasov:2002qd}
N.~A.~Nekrasov,
``Seiberg-Witten prepotential from instanton counting,''
Adv.\ Theor.\ Math.\ Phys.\  \textbf{7}, no.5, 831-864 (2003)
[arXiv:hep-th/0206161 [hep-th]].

\bibitem{Losev:2003py} 
  A.~S.~Losev, A.~Marshakov and N.~A.~Nekrasov,
  ``Small instantons, little strings and free fermions,''
  In *Shifman, M. (ed.) et al.: From fields to strings, vol. 1* 581-621
  [hep-th/0302191].

\bibitem{Nekrasov:2003rj}
N.~Nekrasov and A.~Okounkov,
``Seiberg-Witten theory and random partitions,''
Prog.\ Math.\  \textbf{244}, 525-596 (2006)
[arXiv:hep-th/0306238 [hep-th]].

\bibitem{Baulieu:1997jx}
L.~Baulieu, H.~Kanno and I.~Singer,
``Special quantum field theories in eight-dimensions and other dimensions,''
Commun.\ Math.\ Phys.\  \textbf{194}, 149-175 (1998)
[arXiv:hep-th/9704167 [hep-th]].


\bibitem{Morozov:2018fjb}
A.~Morozov,
``An analogue of Schur functions for the plane partitions,''
Phys.\ Lett.\ B \textbf{785}, 175-183 (2018)
[arXiv:1808.01059 [hep-th]].


\bibitem{Mironov:2019uoy}
A.~Mironov and A.~Morozov,
``On generalized Macdonald polynomials,''
JHEP \textbf{01}, 110 (2020)
[arXiv:1907.05410 [hep-th]].


\bibitem{Nekrasov:2014nea}
N.~Nekrasov and A.~Okounkov,
``Membranes and Sheaves,''
[arXiv:1404.2323 [math.AG]].

\bibitem{Awata:2005fa}
H.~Awata and H.~Kanno,
``Instanton counting, Macdonald functions and the moduli space of D-branes,''
JHEP \textbf{05}, 039 (2005)
[arXiv:hep-th/0502061 [hep-th]].

\bibitem{Iqbal:2007ii}
A.~Iqbal, C.~Kozcaz and C.~Vafa,
``The Refined topological vertex,''
JHEP \textbf{10}, 069 (2009)
[arXiv:hep-th/0701156 [hep-th]].

\bibitem{Awata:2011ce}
H.~Awata, B.~Feigin and J.~Shiraishi,
``Quantum Algebraic Approach to Refined Topological Vertex,''
JHEP \textbf{03}, 041 (2012)
[arXiv:1112.6074 [hep-th]].

\bibitem{Rapcak:2018nsl}
M.~Rapcak, Y.~Soibelman, Y.~Yang and G.~Zhao,
``Cohomological Hall algebras, vertex algebras and instantons,''
[arXiv:1810.10402 [math.QA]].

\bibitem{Eguchi:1992tp} 
  T.~Eguchi, H.~Kanno and S.~K.~Yang,
  ``$W_\infty$ algebra in two-dimensional black hole,''
  Phys.\ Lett.\ B {\bf 298}, 73 (1993)
  [hep-th/9209122].

\bibitem{Eguchi:1993xx} 
  T.~Eguchi, H.~Kanno, Y.~Yamada and S.~K.~Yang,
  ``Topological strings, flat coordinates and gravitational descendants,''
  Phys.\ Lett.\ B {\bf 305}, 235 (1993)
  [hep-th/9302048].


\bibitem{Eguchi:1994np} 
  T.~Eguchi and H.~Kanno,
  ``Toda lattice hierarchy and the topological description of the c = 1 string theory,''
  Phys.\ Lett.\ B {\bf 331}, 330 (1994)
  [hep-th/9404056].


\bibitem{Eguchi:2000fv} 
  T.~Eguchi and H.~Kanno,
  ``Five-dimensional gauge theories and local mirror symmetry,''
  Nucl.\ Phys.\ B {\bf 586}, 331 (2000)
  [hep-th/0005008].
  
  
\bibitem{Eguchi:2003sj} 
  T.~Eguchi and H.~Kanno,
  ``Topological strings and Nekrasov's formulas,''
  JHEP {\bf 0312}, 006 (2003)
  [hep-th/0310235].
  
\bibitem{Eguchi:2003it} 
  T.~Eguchi and H.~Kanno,
  ``Geometric transitions, Chern-Simons gauge theory and Veneziano type amplitudes,''
  Phys.\ Lett.\ B {\bf 585}, 163 (2004)
  [hep-th/0312234].
  
\bibitem{Nakajima} 
H.~Nakajima,
``Lectures on Hilbert schemes of points on surfaces ,''
University Lecture Series, {\bf 18}, AMS (1999).

\bibitem{Nekrasov:JJM} 
  N.~Nekrasov,
  ``Instanton partition function and $M$-theory,''
  Japan. J. Math.  {\bf 4} (2009) 63--93.
 
\bibitem{Jafferis:2007sg}
D.~L.~Jafferis,
``Topological Quiver Matrix Models and Quantum Foam,''
[arXiv:0705.2250 [hep-th]].

\bibitem{Cirafici:2008sn}
M.~Cirafici, A.~Sinkovics and R.~J.~Szabo,
``Cohomological gauge theory, quiver matrix models and Donaldson-Thomas theory,''
Nucl.\ Phys.\ B \textbf{809}, 452-518 (2009)
[arXiv:0803.4188 [hep-th]].

\bibitem{Nekrasov:2017cih} 
  N.~Nekrasov,
  ``Magnificent Four,''
  arXiv:1712.08128 [hep-th].

\bibitem{Nekrasov:2018xsb}
N.~Nekrasov and N.~Piazzalunga,
``Magnificent Four with Colors,''
Commun.\ Math.\ Phys.\  \textbf{372}, no.2, 573-597 (2019)
[arXiv:1808.05206 [hep-th]].


\bibitem{Nekrasov:2015wsu}
N.~Nekrasov,
``BPS/CFT correspondence: non-perturbative Dyson-Schwinger equations and qq-characters,''
JHEP \textbf{03}, 181 (2016)
[arXiv:1512.05388 [hep-th]].

\bibitem{Nekrasov:2016qym} 
  N.~Nekrasov,
  ``BPS/CFT correspondence II: Instantons at crossroads, moduli and compactness theorem,''
  Adv.\ Theor.\ Math.\ Phys.\  {\bf 21}, 503 (2017)
  [arXiv:1608.07272 [hep-th]].

\bibitem{Nekrasov:2016ydq} 
  N.~Nekrasov,
  ``BPS/CFT Correspondence III: Gauge Origami partition function and qq-characters,''
  Commun.\ Math.\ Phys.\  {\bf 358}, no. 3, 863 (2018)
  [arXiv:1701.00189 [hep-th]].

\bibitem{Nekrasov:2016gud} 
  N.~Nekrasov and N.~S.~Prabhakar,
  ``Spiked Instantons from Intersecting D-branes,''
  Nucl.\ Phys.\ B {\bf 914}, 257 (2017)
  [arXiv:1611.03478 [hep-th]].
 
\bibitem{Witten:2000mf}
E.~Witten,
``BPS Bound states of D0 - D6 and D0 - D8 systems in a B field,''
JHEP \textbf{04}, 012 (2002)
[arXiv:hep-th/0012054 [hep-th]].

\bibitem{Ohta:2001dh}
K.~Ohta,
``Supersymmetric D-brane bound states with B field and higher dimensional instantons on noncommutative geometry,''
Phys.\ Rev.\ D \textbf{64}, 046003 (2001)
[arXiv:hep-th/0101082 [hep-th]].

  
\bibitem{Okounkov:2015spn}
A.~Okounkov,
``Lectures on K-theoretic computations in enumerative geometry,''
[arXiv:1512.07363 [math.AG]].


\bibitem{Moore:1997dj} 
  G.~W.~Moore, N.~Nekrasov and S.~Shatashvili,
  ``Integrating over Higgs branches,''
  Commun.\ Math.\ Phys.\  {\bf 209}, 97 (2000)
  [hep-th/9712241].

\bibitem{Moore:1998et} 
  G.~W.~Moore, N.~Nekrasov and S.~Shatashvili,
  ``D particle bound states and generalized instantons,''
  Commun.\ Math.\ Phys.\  {\bf 209}, 77 (2000)
  [hep-th/9803265].
 
 
\bibitem{Carlsson:2013jka} 
  E.~Carlsson, N.~Nekrasov and A.~Okounkov,
  ``Five dimensional gauge theories and vertex operators,''
  Moscow Math.\ J.\  {\bf 14}, no. 1, 39 (2014)
  [arXiv:1308.2465 [math.RT]].

\bibitem{Macdonald} 
I.G.~Macdonald, 
``Symmetric Functions and Hall Polynomials,''  second edn.
Oxford University Press, (1995). 
 
\bibitem{Atiyah:1984tf}
M.~Atiyah and I.~Singer,
``Dirac Operators Coupled to Vector Potentials,''
Proc.\ Nat.\ Acad.\ Sci.\  \textbf{81}, 2597-2600 (1984)

\bibitem{Kanno:1988wm}
H.~Kanno,
``Weil Algebra Structure and Geometrical Meaning of {BRST} Transformation in Topological Quantum Field Theory,''
Z.\ Phys.\ C \textbf{43}, 477 (1989)

  
\bibitem{Shadchin:2004yx}
S.~Shadchin,
``Saddle point equations in Seiberg-Witten theory,''
JHEP \textbf{10}, 033 (2004)
[arXiv:hep-th/0408066 [hep-th]].
 
 
\bibitem{Iqbal:2008ra}
A.~Iqbal, C.~Kozcaz and K.~Shabbir,
``Refined Topological Vertex, Cylindric Partitions and the U(1) Adjoint Theory,''
Nucl.\ Phys.\ B \textbf{838}, 422-457 (2010)
[arXiv:0803.2260 [hep-th]].

\bibitem{Poghossian:2008ge} 
  R.~Poghossian and M.~Samsonyan,
  ``Instantons and the 5D U(1) gauge theory with extra adjoint,''
  J.\ Phys.\ A {\bf 42}, 304024 (2009)
  [arXiv:0804.3564 [hep-th]].

\bibitem{Awata:2009yc}
H.~Awata and H.~Kanno,
``Changing the preferred direction of the refined topological vertex,''
J.\ Geom.\ Phys.\  \textbf{64}, 91-110 (2013)
[arXiv:0903.5383 [hep-th]].


\bibitem{Awata:2009dd} 
  H.~Awata and H.~Kanno,
  ``Quiver Matrix Model and Topological Partition Function in Six Dimensions,''
  JHEP {\bf 0907}, 076 (2009)
  [arXiv:0905.0184 [hep-th]].


%
\bibitem{Benini:2018hjy}
F.~Benini, G.~Bonelli, M.~Poggi and A.~Tanzini,
``Elliptic non-Abelian Donaldson-Thomas invariants of $\mathbb{C}^3$,''
JHEP \textbf{07}, 068 (2019)
[arXiv:1807.08482 [hep-th]].


\bibitem{Fasola:2020hqa}
N.~Fasola, S.~Monavari and A.~T.~Ricolfi,
``Higher rank K-theoretic Donaldson-Thomas theory of points,''
[arXiv:2003.13565 [math.AG]].


\bibitem{Eguchi:1978xp}
T.~Eguchi and A.~J.~Hanson,
``Asymptotically Flat Selfdual Solutions to Euclidean Gravity,''
Phys.\ Lett.\ B \textbf{74}, 249-251 (1978)

\bibitem{Kononov:2019fni}
Y.~Kononov, A.~Okounkov and A.~Osinenko,
``The 2-leg vertex in K-theoretic DT theory,''
[arXiv:1905.01523 [math-ph]].


\end{thebibliography}
\end{document}